\documentstyle[12pt]{JHEP3}

\newcommand{\be}{\begin{equation}}
\newcommand{\bea}{\begin{eqnarray}}
\newcommand{\ee}{\end{equation}}
\newcommand{\eea}{\end{eqnarray}}

\def\theequation{\arabic{section}.\arabic{equation}}
\newcommand{\bn}[2]{\left(\begin{array}{c} #1\\ #2
\end{array}\right)}

\title{ \bf Higher Spin Gauge Fields Interacting with Scalars: The
Lagrangian Cubic Vertex} 

\author{\bf Angelos
Fotopoulos\\Dipartimento di Fisica Teorica dell'Universit\`a di Torino
and INFN\\Sezione di Torino,
via P. Giuria 1, I-10125 Torino, Italy}
\vspace{2cm}

\author{\bf Nikos Irges, Anastasios C.
Petkou and  Mirian Tsulaia\\
Department of
Physics, University of Crete, 71003 Heraklion, Crete, Greece}

\vspace{3cm}

\abstract{We apply a recently presented
BRST procedure to construct the Largangian cubic vertex of higher-spin
gauge field triplets interacting with massive free scalars. In
flat space, the spin-$s$ triplet propagates the series of irreducible
spin-$s$, $s-2$,..,0/1 modes which couple independently to
corresponding conserved currents constructed from the scalars. The
simple covariantization of the flat space results is not enough in
AdS, as new interaction vertices appear.  We present in detail the
cases of spin-2 and spin-3 triplets coupled to scalars. Restricting to a
single irreducible spin-$s$ mode we uncover previously obtained results.
We also present an alternative derivation of the lower spin results based
on the idea that higher-spin  gauge fields arise from the gauging of higher
derivative symmetries of free matter Lagrangians. Our results can be readily
applied to holographic studies of higher-spin gauge theories.}
\begin{document}


\vfill



\section{Introduction}

Constructing consistent interactions for higher-spin (HS) gauge
fields  is an old problem (see \cite{Vasiliev:2004qz} for recent
reviews). A crucial step for its resolution in AdS spaces was taken
some years ago by Fradkin and Vasiliev \cite{Fradkin:1986qy} (see
also \cite{Vasiliev:1990en}). Since then, the study of HS gauge
theories has enjoyed a remarkable renaissance and a wealth of new
and interesting results have appeared
\cite{Sezgin:2005pv}--\cite{Benincasa:2007xk} (see also
\cite{Fronsdal:1978rb}-- \cite{Ouvry:1986dv} for the earlier work).

One of the approaches to the interaction of HS  gauge fields is
based on the BRST-like methods  ( see e.g., \cite{Bekaert:2003uc}
for a review). This is particularly appealing as it resembles
similar constructions in string field theory \cite{Neveu:1986mv}--
\cite{Gross:1986ia}. In particular a model of interacting massless
HS gauge fields can be obtained using a cubic vertex of the open
string theory \cite{Fotopoulos:2007nm}.
 However in the general case of
 interacting massless HS fields there is no analog
of overlap conditions such as that is present in the Open String
Field Theory and therefore one has to consider a general polynomial
of the corresponding matter and ghost oscillators.

In \cite{Buchbinder:2006eq} a systematic method for the construction
of the general cubic coupling of any three HS gauge fields in flat
and AdS spaces based on the triplet construction
\cite{Francia:2002pt}--\cite{Barnich:2005bn} was presented. In the
present note we apply our method to the simplest case; the
interaction between one HS triplet and two massive free scalars.
Despite its apparent simplicity this is still a highly non-trivial
case since it requires the construction of an infinite number of
conserved currents, made out of scalars,  that couple properly to HS
gauge fields. This is also an important case since, as we will see
in Section 7, it elucidates the emergence of HS gauge
 fields via the gauging of higher derivative symmetries of free matter Lagrangians
 (see \cite{Bekaert:2007mi} for the discussion about self-adjoint operators
 in the gauging of HS symmetries).
 In a holographic setting, our results imply the
 existence of an infinite set of Ward
 identities involving scalar operators in
 boundary CFTs that are dual to HS gauge theories in AdS spaces.

The paper is organized as follows:

In Section 2 we briefly review our general method of constructing
free and interacting Lagrangians for HS gauge fields
\cite{Buchbinder:2006eq}--\cite{Sagnotti:2003qa}. In Section 3 we
present the general result for the Lagrangian cubic interaction
vertices between one triplet and two massive free scalars in flat
space. We note the emergence of a pattern; the irreducible
spin-$s,s-2,...,0/1$ modes propagated by the spin-$s$ triplet couple
independently to corresponding conserved currents constructed from
the scalars. In Section 4 we outline how we obtain the general
result for AdS. The simple flat space pattern is no more valid and
new interaction vertices appear at order $1/L^2$. To keep our
presentation clear we relegate the explicit lengthy expressions in
the Appendix. In Section 5 we present the explicit formulae for the
spin-2 and spin-3 cases in flat and AdS spaces. In Section 6 we
present the results for the cubic vertex of irreducible HS gauge
fields interacting with massive free scalars and show that our
formulae reproduce known past results
\cite{Anselmi:1999bb}--\cite{Manvelyan:2004mb}.  Extensive discussions of conformal HS currents were presented also in 
\cite{Vasiliev:1999ba,Konstein:2000bi,Prokushkin:1999xq}. In Section 7
we re-derive the spin-2 and spin-3 vertices in flat and AdS spaces
by an alternative method based on the idea that HS gauge fields
arise from the gauging of high derivative symmetries of free matter
Lagrangians \cite{Mikhailov:2002bp}--\cite{Eastwood:2002su}. This
procedure opens up the possibility for explicit study of the HS
gauge symmetry acting on scalars, however we leave this interesting
idea for a future work. Section 8 contains a summary and the outlook
of our work. Important notation and some lengthy formulae appear in
the five Appendices.

\setcounter{equation}0\section{The BRST approach to the HS cubic vertex}\label{BE}

In this Section we review the BRST approach to constructing the
cubic interaction of HS gauge fields. More details can be found in
\cite{Sagnotti:2003qa}, \cite{Buchbinder:2006eq}. We consider an AdS
space with  radius $R_{AdS}=L$; the corresponding equations for flat
space are simply obtained by setting the AdS curvature to zero.

The full interacting Lagrangian can be written as
\cite{Neveu:1986mv} -- \cite{Gross:1986ia}
\begin{equation} \label {LIBRST}
{L} \ = \ \sum_{i=1,2,3} \int d c_0^i \langle \Phi_i |\, Q_i \,
|\Phi_i \rangle \ + g( \int dc_0^1 dc_0^2  dc_0^3 \langle \Phi_1|
\langle \Phi_2|\langle \Phi_3||V \rangle + h.c)\,, \end{equation}
where $|V\rangle$ is the cubic vertex and $g$ is a dimensionless
coupling constant.\footnote{Each term in the Lagrangian
(\ref{LIBRST}) has length dimension $- {\cal D}$. This requirement
holds true for each space-time vertex contained in (\ref{LIBRST})
after multiplication by an appropriate power of the length scale of
the theory, as discussed in \cite{Buchbinder:2006eq}.}
 To describe the cubic interaction of HS gauge fields we
 introduced three vectors $|\Phi_i \rangle$ $(i=1,2,3)$ associated to  three -
generally different - Fock spaces spanned by the oscillators
\begin{equation} \label{3HIL}
[\alpha_\mu^i, \alpha_\nu^{j,+} ] = \delta^{ij} g_{\mu \nu}, \quad
\{ c^{i,+}, b^j \} = \{ c^i, b^{j,+} \} = \{ c_0^i , b_0^j \} =
\delta^{ij}\ .
\end{equation}
The vacuum  in each one of the Fock spaces is defined as
\begin{equation}
c|0\rangle\ = \ 0 \ , \qquad b|0\rangle\ = \ 0 \ , \qquad
b_0|0\rangle\ =\ 0 \ ,\qquad \alpha^\mu|0 \rangle =0\,.
\end{equation}
Each of the fields $|\Phi_i \rangle$ (so called ''triplets") has the
form
\begin{equation}\label{Phifield}
|\Phi  \rangle = |\phi \rangle + c^+ \ b^+\ |D\rangle + c_0\ b^+\
|C\rangle\,,
\end{equation}
with
\begin{equation}
 |\phi \rangle = \frac{1}{s!}\, h_{\mu_1 \ldots \mu_s}(x)
\alpha^{\mu_1 +} \ldots \alpha^{\mu_s +}\; |0\rangle,
\end{equation}
\begin{equation}
  |D\rangle = \ \frac{1}{(s-2)!}\, D_{\mu_1 \ldots \mu_{s-2}}(x)
\alpha^{\mu_1 +} \ldots \alpha^{ \mu_{s-2} +} \,  \; |0\rangle \ ,
\end{equation}
\begin{equation}
 |C \rangle \ = \ \frac{-i}{(s-1)!}\, C_{\mu_1 \ldots
\mu_{s-1}}(x) \alpha^{\mu_1 +}  \ldots \alpha^{ \mu_{s-1} +} \,
 \; |0\rangle
\end{equation}
and the identically nilpotent BRST charge
 \cite{Sagnotti:2003qa} in each of the Fock spaces
has the form
 \begin{eqnarray} \label{brst}
Q& =&c_0\Bigl( l_0 \, + \, \frac{1}{L^2} (  N^2 - 6N +6 +{\cal D} \,
- \, \frac{{\cal D}^2}{4} - 4M^+M +
 c^+b(4N-6) \nonumber\\
&& +b^+c (4N-6) +12 c^+b b^+c
-8 c^+b^+M  +8 M^+bc)\Bigl)\nonumber \\
&&+ c^+l + cl^+ - c^+cb_0\,
\end{eqnarray}
with
\begin{equation} \label{lapl} l_0 \ = \
g^{\mu \nu}  p_\mu p_\nu\,, \quad l= \alpha^\mu p_\mu, \quad l^+ =
\alpha^{\mu +} p_\mu \,,
\end{equation}
and
\begin{equation} N \ = \ \alpha^{\mu +}  \alpha_{\mu} \ + \
\frac{\cal D}{2} , \quad M= \frac{1}{2} \alpha^\mu \alpha_\mu
\, .\label{Nop}
\end{equation}
The momentum operator $p_\mu$ is defined as
\cite{Buchbinder:2006ge}
 \begin{equation}
\label{pop}
p_\mu \ = \  -\; i \, \left(
\nabla_\mu + \omega_{\mu}^{ab} \, \alpha_{\; a}^+\,
  \alpha_{ \; b} \right) \, ,\,\,\,\,\\ \alpha^a=e^{a}_\mu\alpha^\mu\,,
\end{equation}
where    $e^a_\mu$ and $\omega_\mu^{ab}$ are the vierbein and   the spin connection of AdS and
$\nabla_\mu$ is the AdS covariant derivative. The Lagrangian is
invariant up to terms of order $ g^2$ under the nonabelian gauge
transformations
\begin{equation}\label{BRSTIGT1}
\delta | \Phi_{i} \rangle  =  Q_{i} | \Lambda_{i} \rangle  - g \int
dc_0^{i+1} dc_0^{i+2}[(  \langle \Phi_{i+1}|\langle \Lambda_{i+2}|
+\langle \Phi_{i+2}|\langle \Lambda_{i+1}|) |V \rangle] + O(g^2)\,,
\end{equation}
with  gauge transformation parameters
\begin{equation}\label{GPdef}
|\Lambda_i \rangle = b^{i,+} |\lambda^i \rangle=\frac{i}{(s-1)!}
\lambda^i_{\mu_1 \mu_2... \mu_{s-1}}(x) \alpha^{i,\mu_1  +}
\alpha^{i,\mu_2 +}... \alpha^{i,\mu_{s-1} +} b^{+} |0 \rangle \,,
\end{equation}
provided that the vertex satisfies the BRST invariance condition
\begin{equation}\label{VBRST}
 \tilde Q|V \rangle=0\, \ , \quad \tilde Q =\sum_i Q_i .
\end{equation}
The vertex operator $|V \rangle$ has  ghost number $+3$ and its structure is
\begin{equation}\label{Vertex1}
| V \rangle= V |- \rangle_{123}, \quad |-\rangle_{123}= c^1_0 c^2_0
c^3_0\ |0 \rangle_{1} \otimes |0 \rangle_{2} \otimes |0
\rangle_{3}\,,
\end{equation}
where $V$ is an unknown function of
the rest of the oscillators with ghost number zero. Note that  equation (\ref{VBRST}) determines
the
 vertex   up to $\tilde{Q}$-exact terms
\begin{equation}\label{VFR}
\delta |V \rangle= \tilde{Q} |W \rangle\,,
\end{equation}
where  $|W \rangle$  is an operator with total ghost number $+2$.
The BRST-exact terms correspond to the ones which can be obtained by
field redefinitions from the free Lagrangian.
 To simplify
the analysis of equation (\ref{VBRST}) we  define the operator
\begin{equation}\label{DefN}
\tilde N=\alpha^{\mu, i+} \alpha^i_{\mu}+b^{i,+} c^{i}+
c^{i,+}b^i\,.
\end{equation}
This operator commutes with the BRST charges $Q_i$.
 This means
 that the vertex can be expanded in a sum of terms, each  with fixed eigenvalues $K$
of the operator $\tilde N$ as
\begin{equation}
|V\rangle =\sum_{K}|V(K)\rangle\,.
\end{equation}
Therefore,  equation (\ref{VBRST}) can be split into the set of
algebraic equations
\begin{equation}\label{VBRST1}
\sum_i Q_i V(K) =0
\end{equation}
for each value of $K$. The systematic solution of these equations
determines the type the  of cubic vertex.

\setcounter{equation}0\section{The cubic vertex in flat
space}\label{FV}

In this section we will use our general approach to construct  the
cubic coupling of an HS triplet with two scalars in flat space. The
corresponding coupling for irreducible HS fields has been discussed
\cite{Berends:1985xx} and more recently  in \cite{Anselmi:1999bb,
Manvelyan:2004mb}. The merit of considering the triplet is twofold.
Firstly, our general BRST construction is systematic, can be
straightforwardly generalized to $AdS_{{\cal D}}$ and is relevant
for string field theory constructions. Secondly, our construction
gives rise to the
 idea that HS gauge fields arise from the gauging of higher
 derivative symmetries of free matter Lagrangians.

To proceed, we put  the triplet of spin $s$ in Fock space 3 and the
two  scalars $\phi_1$ and $\phi_2$ into  Fock spaces $1$ and
$2$. Therefore, the matter and complex ghost
oscillators appear only in Fock space $3$. This allows us to write down the most general polynomial in
terms of the expansion in ghost variables
as (see Appendix A for the definition of
  the various operators used below)
\begin{equation}\label{Vs00F}
\langle V|= \ _{123}\langle -|\Bigl\{X^1 + X^2_{33} \gamma^{33}+
X^3_{3j} \beta^{3j}\Bigl\}\,,\,\,\, j=1,2,3\,,
\end{equation}
where
\begin{equation}\label{Xs00F}
X^{1}= X^{1}_{n_1,n_2,n_3;m_3,k_3;p_3,}\,\,
(l_0^1)^{n_1}(l_0^2)^{n_2}(l_0^2)^{n_3} (l^{3})^{m_3} (I^{3})^{k_3}
(M^{33})^{p_3}\,,
\end{equation}
and with the same expansions for  the coefficients $X^2_{33}$ and
$X^3_{3j}$.
 We will apply the field redefinition (FR) scheme outlined in
\cite{Buchbinder:2006eq} in order to eliminate the $l_0^{i}$
from all three matrix elements, and the $l^3$ dependence from
$X^1$ and $X^2$. This amounts to removing  the
indices $n_1,n_2,n_3$ from the expressions of $X^1$, $X^2_{33}$  and $X^3_{3j}$.
 Then, the  BRST invariance condition (\ref{VBRST}) simplifies to
\begin{eqnarray}\label{Es00A1}
&&\hspace{-1cm}_{123}\langle -| (I_3)^{k_3} \ (l_{33})^{m_3} \ (M_{33})^{p_3} \Bigl[
-X^3_{31; k_3,m_3,p_3} \ l_0^{11}  - X^3_{32; k_3,n_3,p_3} \
l_0^{22}-  \nonumber \\
&& \hspace{1cm}X^3_{33; k_3,n_3,p_3}l_0^{33} + \delta_{m_3,0} (l^{+}_{33} \
X^1_{k_3,0,p_3}- l_{33}\ X^2_{33; k_3, 0, p_3})\Bigl]=0\,.
\end{eqnarray}
We can solve (\ref{Es00A1})  to derive
\begin{eqnarray}\label{Es00F}
X^3_{31;0,k_3-1,p_3}&=& k_3 X^1_{0,k_3,p_3}\,, \nonumber \\
X^3_{32;0,k_3-1,p_3}&=& -k_3 X^1_{0,k_3,p_3}\,, \nonumber \\
X^2_{33;0,k_3,p_3-1}&=& -p_3 X^1_{0,k_3,p_3} \,,\nonumber \\
X^3_{33;m_3,k_3,p_3}&=& 0 \,,\nonumber \\
X^3_{ij; m_3\neq 0 ;k_3,p_3}&=& X^3_{33; m_3 \neq
0;k_3,p_3}=X^1_{m_3\neq 0;k_3,p_3}=0\,.
\end{eqnarray}
For a triplet of spin-$s$ the solution takes the form (we drop the
$m_3$ index from now on):
\begin{eqnarray}\label{Vs00F2}
&\langle V|= \ _{123}\langle -|\Bigl\{X^1_{k_3,p_3} -
(p_3+1)X^1_{k_3,p_3+1}
\gamma^{33,+} \nonumber \\
&+(k_3+1)X^1_{k_3+1,p_3} \beta^{31,+}- (k_3+1)X^1_{k_3+1,p_3}
\beta^{32,+}\Bigl\} (I^{+,3})^{k_3} (M^{+,33})^{p_3}\,,
\end{eqnarray}
with
\begin{equation}\label{spk}
s=2p_3+k_3, \quad X^1_{k_3,p_3}= 2^p_3\ C_{s,p_3}\quad
k_3=0,1,..,s\,.
\end{equation}
 Using the ''momentum conservation" condition
$p_\mu^1 + p_\mu^2 + p_\mu^3=0$ and the solution above we can write
the gauge transformation rules for scalars (\ref{BRSTIGT1})
 in the form
\begin{eqnarray}\label{FGT2}
&&\delta \phi_a = g \   \sum^{[{s-1 \over 2}]}_{p=0} \ \xi^{ab}
\ (s-2p) \ C_{s,p} \sum_{r=0}^{s-1-2p} \ \bn{s-1-2p}{r} \ 2^r\nonumber \\
&&(\partial^{\mu_{r+1}} \dots
\partial^{\mu_{s-1-2p}}\lambda^{[p]}_{\mu_{1} \dots \mu_{s-1-2p}})  \
(\partial^{\mu_{1}} \dots
\partial^{\mu_r}\phi_b)
\end{eqnarray}
 \begin{equation}\label{xi}
 \xi^{12}=(-1)^{s-1}\xi^{21}=-1, \ \xi^{11}=\xi^{22}=0 \quad a,b =1,2\,,
 \end{equation}
where $C_{s,p}$ are arbitrary parameters. Note that the gauge transformations of the scalars are
nonabelian, whereas the gauge transformations of the fields
in the spin-$s$ triplet remain abelian
\begin{equation}
\delta h_{\mu_1 \dots \mu_s} = \partial_{\{\mu_1} \lambda_{\mu_2 \dots
\mu_s \}}, \quad \delta C_{\mu_1 \dots \mu_{s-1}} = \Box \lambda_{\mu_1
\dots \mu_s}, \quad \delta D_{\mu_1 \dots \mu_{s-2}} = \partial^\mu
\lambda_{\mu, \mu_1 \dots \mu_{s-2}}\,..
\end{equation}
 At this point we should
emphasize that we have not imposed any symmetry between the scalars
$\phi_a $. It is obvious though from (\ref{xi}) that for HS triplets
with $s$ odd ($s=2k+1$) one needs at least two {\it different} real
scalars (or alternatively one complex scalar) to have a non-zero
coupling  \footnote{Indeed setting $\phi_1=\phi_2$ in (\ref{FGT2}) and taking
 into account (\ref{xi}) leads into an inconsitency for s-odd while it is
 allowed for s-even.}. The simple example of this is the $s=1$ case where we find
linearized scalar electrodynamics. For even $s$ HS triplets $(s=2k)$
the condition (\ref{xi}) gives no obstruction to  coupling with a
single real scalar.

Finally, after some rather lengthy rearrangement we obtain the cubic
interaction terms in the Lagrangian as
\begin{eqnarray}\label{Ls00F}
{\cal L}_{s00}&=& \int d^d x \ \Bigl\{\sum^{[{s \over 2}]}_{p=0} \
C_{s,p} \ {\cal W}^{[p]}_{\mu_1 \dots \mu_{s-2p}} \times
\nonumber \\
 &&\sum_{r=0}^{s-2p} \ \bn{s-2p}{r} \ (-1)^r
\ (\partial^{\mu_1} \dots
\partial^{\mu_r}\phi_1)\ (\partial^{\mu_{r+1}} \dots
\partial^{\mu_{s-2p}}\phi_2) \ + \ h.c. \Bigl\} \nonumber \\
&=& \int d^d x \ \sum^{[{s \over 2}]}_{p=0} \ C_{s,p} \ {\cal
W}_{p} \cdot J_{s-2p}\ + \ h.c.\,,
\end{eqnarray}
where we have used the binomial coefficients $\bn{n}{m} $ and $[{s \over 2}]$ is the
integer part of ${s\over 2}$. ${\cal
W}_p$ is defined in \cite{Francia:2002pt}
\begin{equation}\label{hatS}
{\cal W}_{p}= h^{[p]}-2p \ D^{[p-1]}\,, \,\,\,\delta {\cal W}_{p}= \partial \Lambda^{[p]} \,,\nonumber
\end{equation}
and
\begin{equation}\label{Jp}
J_{s-2p}=\sum_{r=0}^{s-2p} \ \bn{s-2p}{r} \ (-1)^r \
(\partial^{\mu_1} \dots
\partial^{\mu_r}\phi_1)\ (\partial^{\mu_{r+1}} \dots
\partial^{\mu_{s-2p}}\phi_2)\,.
\end{equation}
The fields ${\cal W}_p$ define a chain of lower spin fields
contained in the triplet as  can easily be seen from their gauge
transformation properties  (\ref{hatS}). They are rank $s-2p$
symmetric tensors. Hence, the  currents of (\ref{Jp}) are also
symmetrized and by virtue of the transformation properties
(\ref{hatS}) are conserved. We see a pattern emerging: given a
general free scalar Lagrangian we can construct the series of
spin-$s,s-2,...,0/1$ symmetric conserved currents (\ref{Jp}) that
couple properly to a spin-$s$ triplet. In Section 7 we will try to
understand the deep reason for the existence of such currents.

\setcounter{equation}0\section{The cubic vertex
 in AdS}\label{VAdS}

In this section we present the construction of the cubic vertex of a
triplet  coupled to two massive free scalars in AdS space. In this
case the calculations are rather more involved, nevertheless we will
be able to obtain a relatively simple result for the vertex.

The vertex still has the form (\ref{Vs00F}). Next, we choose an FR
(field redefinition) scheme where one  can eliminate all $l_0^{ii}$
dependence in (\ref{XEXP}) and set $X^3_{33}=0$  in (\ref{Vs00F}).
With this choice we are able to eliminate any $l^{i}$ dependence
from $X^2_{33}$. Therefore one has  the expansion of the vertex
\begin{eqnarray}\label{Vs00A}
 \langle V |&= &\ _{123}\langle -| (I_3)^{k_3} \ (l_{33})^{m_3} \
(M_{33})^{p_3} \Bigl\{ X^1_{k_3, m_3, p_3} + X^2_{33; k_3, 0, p_3} \
\gamma^{33}
\nonumber \\
&+& X^3_{31; k_3,m_3,p_3} \ \beta^{13}+ X^3_{32; k_3,m_3,p_3} \
\beta^{23}\Bigl\}\,.
\end{eqnarray}
The BRST invariance gives the equation
\begin{eqnarray}\label{Es00A}
&&_{123}\langle -| (I_3)^{k_3} \ (l_{33})^{m_3} \ (M_{33})^{p_3}
\Bigl\{ -X^3_{31; k_3,m_3,p_3} \ (l_0^{11} - {2{\cal D} -6 \over
L^2}) - \\ \nonumber && X^3_{32; k_3,m_3,p_3} (l_0^{22} -{2{\cal D}
-6 \over L^2})+ l^{+}_{33} X^1_{k_3,m_3,p_3}- l_{33}\delta_{m_3,0}
X^2_{33; k_3, 0, p_3}\Bigl\}=0\,.
\end{eqnarray}
In order to arrive at the equivalent of (\ref{Es00F}) we will have
to commute all creation operators $\alpha^{+, 3}_\mu$ to the left
but  we will also have to eliminate one of the three momenta i.e.,
$p^{\mu, 3}$ using ''momentum conservation". In flat space
commutativity of momenta makes this a very easy task. In AdS this
becomes rather involved due to the relation (\ref{Dcom}) (see also
\cite{Buchbinder:2006eq}).
 The rules one should apply are the following\footnote{We set
$L^2=1$ in what follows and restore it at the end by dimensional
analysis.}:
\begin{itemize}
\item In order to use "momentum conservation"  we move
 the operators $p^{\mu}_{ 3}$ to the far left of the expression.
 Then we substitute $p^\mu_1+p^\mu_2+p^\mu_3=0$. For example, instead of writing
\begin{equation}
(l_{32})  p_{\rho,3}  (l_{32}) p_2^{\rho}=p_{\rho,3}(l_{32})
(l_{32}) p_2^{\rho}\,,
\end{equation}
which translates to
\begin{equation}
 \int   d^D x \sqrt{-g}  (\nabla^\rho \Lambda^{\mu \nu})
  (\nabla_{\mu} \nabla_{\nu} \nabla_\rho \phi^2)  \phi^1\,
\end{equation}
we use\begin{equation} - (p^1_{\mu,}+p_\mu^2)(l_{32})  (l_{32})
p_2^{\mu}\,,
\end{equation}
which translates
into
\begin{equation} \label{examp1}
 -\int   d^D x \sqrt{-g}   \Lambda^{\mu \nu }
 [ ( \nabla_{\mu} \nabla_{\nu} \nabla^\rho  \phi^2) (
\nabla_{\rho} \phi^1)+  (\nabla^\rho \nabla_\mu \nabla_{\nu}
\nabla_\rho \phi^2) \phi^1]\,,.
\end{equation}

\item We will then use the equations of Appendix \ref{ApC} to move
any ''non-contracted" momenta $p_\mu^i, \ i=1,2$ to the right until
they  form operators $l_{0}^{11}, l_0^{22}, l^{31}$ or $l^{32}$ with
operators $p_{\mu}^1, \ p_{\mu}^2$ or $\alpha_\mu^{3}$. In the
present example we should push the operators $p_\mu^1$ and $p_\mu^2$
to the right until they form the operators $l_0^{11}$ and $l_0^{12}$
when combined with  $p^\mu_{2}$. This process will generate
 terms proportional to $\frac{1}{L^2}$. For the example above they are
\begin{equation}
-\int   d^D x \sqrt{-g}\frac{1}{L^2}[\Lambda^\mu_\mu (\Box \phi^2)
\phi^1 + (1- 2{\cal D}) \Lambda^{\mu \nu} (\nabla_\mu \nabla_\nu
\phi^2)\phi^1 ]
\end{equation}
  which can be seen from the
 second term in (\ref{examp1}) when pushing the covariant derivative
 $\nabla_\rho$ to the right.

\item We will commute creation oscillators to the left. In doing so
we will once more generate non-contacted momenta $p^{\mu,i} \ i =
1,2$ which in turn have to be pushed to the right and will generate
further $1/L^2$ terms as  explained in the previous step.

\item Finally the ordering  rule is that all operators
 $l_0^{11}, l_0^{22}$
 which do not commute
with $I^3,$ and  $l^3$, are to be brought to the extreme right of
the equation so that we compare operator expressions which have the
same ordering.

\end{itemize}

This  procedure  results in some quite lengthy equations but  our
choice of FR scheme which  has eliminated $X^3_{33}$ simplifies the
problem. Actually, we have to perform the manipulations described
above only for the third and fourth terms in  (\ref{Es00A}).
 For the fourth terms we just push the operator $p^{\mu, 3}$ to the
  left, then use
  ''momentum conservation" and then push the operator $p^{\mu,1}+p^{\mu,2}$ to the
right as described above. The third term is the hardest one  since
it requires  performance of commutators both among momenta and
among oscillators.

The solution for generic triplet is quite involved but
straightforward. In the Appendices we give
some explicit formulae that are used  in the manipulations described above. The full solution
for the triplet will not be presented here. Instead, as an
illustration of our technique we present the two simplest examples
describing the interaction of spin-$2$ and spin-$3$ triplets with
two massive free scalars.

\setcounter{equation}0\section{Explicit examples}
\subsection{Spin-$2$ with two scalars}

 Since the oscillators
$\alpha_\mu^{i,+}, c^{i,+}$ and $b^{i,+}$ occur only in the third
Fock space we omit the index $i$ for them in what follows. The field will will using are
\begin{equation} \label{phi12}
|\Phi_1\rangle = \phi_1(x)|0\rangle, \quad |\Phi_2\rangle =
\phi_2(x)|0\rangle,
\end{equation}
\begin{equation}
|\Phi_3\rangle = (\frac{1}{2!} h_{\mu \nu}(x) \alpha^{\mu +}
\alpha^{\nu +} + D(x)c^+ b^+ - i C_\mu(x) \alpha^{\mu +} c_0^3
b^+) |0 \rangle\,,
\end{equation}
\begin{equation}
|\Lambda \rangle = i\lambda_\mu (x) \alpha^{\mu +} b^+ |0 \rangle\,.
\end{equation}
The Lagrangian has the form
\begin{equation}\label{le}
L= L_{free} + L_{int}\,,
\end{equation}
\begin{eqnarray}\label{lf200}
L_{free}&=& (\partial_\mu \phi_1) (\partial^\mu \phi_1)
+(\partial_\mu \phi_2) (\partial^\mu \phi_2) +m^2(\phi_1^2+\phi_2^2)+ (\partial_\rho h_{\mu
\nu}) (\partial^\rho h^{\mu \nu})\nonumber\\
&&-4 (\partial_\mu h^{\mu \nu})
C_\nu - 4 (\partial_\mu C^\mu) D -2(\partial_\mu D)
(\partial^\mu D) + 2C_\mu C^\mu\,,
\end{eqnarray}
\begin{eqnarray}\label{li200}
L_{int}&=& C_{2,0} \ (  h^{\mu \nu}(\partial_{\mu}
\partial_{\nu} \phi_1)\phi_2 + h^{\mu
\nu}(\partial_{\mu}
\partial_{\nu} \phi_2)\phi_1
-  2h^{\mu \nu}(\partial_{\mu}  \phi_1)(\partial_{\nu} \phi_2)) \nonumber \\
 && - C_{2,1} \ \phi_1 \phi_2 (h^\mu_\mu -2D)\,.
\end{eqnarray}
The relevant gauge transformations are
\begin{equation}
\label{gt21}
\delta \phi_1 = C_{2,0} \ (2 \lambda^\mu \partial_\mu \phi_2 +
\phi_2
\partial_\mu \lambda^\mu)\,,
\end{equation}
\begin{equation}\label{gt22}
\delta \phi_2 = C_{2,0} \ (2 \lambda^\mu \partial_\mu \phi_1 +
\phi_1
\partial_\mu \lambda^\mu)\,,
\end{equation}
\begin{equation}\label{g200}
\delta h_{\mu \nu} =  \partial_\mu \lambda_\nu +\partial_\nu
\lambda_\mu, \quad \delta C_\mu  = \Box \lambda_\mu, \quad \delta D
=\partial_\mu \lambda^\mu\,.
\end{equation}
According to our general construction, given in the section 2 we have
 obtained the cubic vertex which involves two different scalars and the
 triplet with higher spin 2.
To obtain the interaction of a single scalar with the spin-2 field we need to set  $\phi_1=\phi_2$ \footnote{Notice that setting i.e. $\phi_2=0$ is meaningless since in our formalism that would mean to consider two Hilbert spaces, hence no cubic interaction vertex.}. 
It should also be noticed that for $\phi_1=\phi_2$ (\ref{li200}) is equivalent to the linearized interaction of a scalar field with gravity as we explain in  Section 7.1. and in particular in equations (7.4) and (7.6). 
 The generalization for the coupling of a spin-2 triplet with an arbitrary
 number of scalar fields $n$ goes in an analogous manner with the constants
 $C_{2,0}$ becoming $n \times n$ matrices. Similar things apply to the
 couplings of scalars with any HS triplet. For simplicity in what follows we
 will discuss only the two scalar case which the reader can generalize
 easily to the $n$ scalar case.

In $AdS_{{\cal D}}$ we replace ordinary with covariant derivatives.
There will be no other changes for the gauge transformation rules
(i.e., for all fields $\delta_{AdS}= \delta$) (\ref{g200}) except
for
\begin{equation}
\delta_{AdS}C_\mu = \delta C_\mu + \frac{1-{\cal D}}{L^2}\lambda_\mu\,,
\end{equation}
The free Lagrangian is modified to include the standard AdS "mass -terms" of order $1/L^2$ 
\begin{equation}
\Delta L_{free}= -\frac{1}{L^2} (2h_\mu^\mu h_\nu^\nu - 16 h_\mu^\mu
D + 2h_{\mu \nu} h^{\mu \nu} + (4 {\cal D} +12) D^2 + (2{\cal D}-6)\
( \phi_1^2+\phi_2^2))\,.
\end{equation}
The interaction part also changes and gets an additional piece
\begin{equation}
\label{lintAdS200}
\Delta L_{int.} = C_{2,0}\  \frac{{\cal D}-1}{L^2} D \phi_1 \phi_2\,.
\end{equation}
This is an additional interaction of the $D$ scalar with a "spin-0" current.

\subsection{Spin-$3$ with two scalars}

The spin-3 triplet is described by the field
\begin{equation}
|\Phi_3\rangle = (\frac{1}{3!} h_{\mu \nu \rho}(x) \alpha^{\mu +}
 \alpha^{\nu +}\alpha^{\rho +} + D_\mu(x) \alpha^{\mu +} c^+ b^+ -
 \frac{i}{2} C_{\mu \nu}(x) \alpha^{\mu +}\alpha^{\nu +} c_0^3 b^+)
|0 \rangle,
\end{equation}
\begin{equation}
|\Lambda \rangle = \frac{i}{2}\lambda_{\mu \nu} (x) \alpha^{\mu +}
\alpha^{\nu +} b^+ |0 \rangle\,.
\end{equation}
The relevant scalar and gauge transformations are
\begin{equation}
\label{gt31}
\delta \phi_1 = 3i \ C_{3,0}\ (4 \lambda^{\mu \nu} \partial_\mu
\partial_\nu \phi_2 + \phi_2
\partial_\mu \partial_\nu\lambda^{\mu \nu} +
4 (\partial_\mu \phi_2)(\partial_\nu \lambda^{\mu \nu} )) + i \
C_{3,1}\ \phi_2 \lambda_\mu^\mu\,,
\end{equation}
\begin{equation}\label{gt32}
\delta \phi_2 = -3i \ C_{3,0}\ (4 \lambda^{\mu \nu}
\partial_\mu
\partial_\nu \phi_1 + \phi_1
\partial_\mu \partial_\nu\lambda^{\mu \nu} +
4 (\partial_\mu \phi_1)(\partial_\nu \lambda^{\mu \nu} )) - i\
C_{3,1}\  \phi_1 \lambda_\mu^\mu\,,
\end{equation}
\begin{equation}\label{g300}
\delta h_{\mu \nu \rho} =  \partial_\mu \lambda_{\nu \rho}
+\partial_\nu \lambda_{\mu \rho} +\partial_\rho \lambda_{\mu \nu} ,
\quad \delta C_{\mu \nu}  = \Box \lambda_{\mu \nu}, \quad \delta
D_\mu =\partial_\nu \lambda^\nu_\mu\,.
\end{equation}
The free and interacting parts of the Lagrangian have the form
\begin{eqnarray}
L_{free}&=& (\partial_\mu \phi_1) (\partial^\mu \phi_1)+
(\partial_\mu \phi_2) (\partial^\mu \phi_2) +m^2(\phi_1^2+\phi_2^2)+  (\partial_\tau h_{\mu
\nu \rho})(\partial^\tau h^{\mu \nu \rho})
\\ \nonumber
&&- 6 (\partial_\rho h^{\mu \nu \rho}) C_{\mu \rho} -12(\partial_\mu
C^{\mu \nu})D_\nu - 6 (\partial_\mu D_\nu) (\partial^\mu D^\nu) +
3C_\mu C^\mu\,,
\end{eqnarray}
\begin{eqnarray} \label{lint300}\nonumber
L_{int.}&=&i\  C_{3,0}\ ( h^{\mu \nu \rho} \phi_1
\partial_\mu
\partial_\nu \partial_\rho \phi_2 -
h^{\mu \nu \rho} \phi_2 \partial_\mu
\partial_\nu \partial_\rho \phi_1 -3
h^{\mu \nu \rho} (\partial_\mu \partial_\nu \phi_2)(\partial_\rho
\phi_1) \\ \nonumber && +  3 h^{\mu \nu \rho} (\partial_\mu
\partial_\nu
\phi_1)(\partial_\rho \phi_2)) \\
&&   +i \ C_{3,1}\ (h^{\mu \nu}_\nu -2D^\mu) (\phi_1
\partial_\mu \phi_2 -\phi_2
\partial_\mu \phi_1  ) + h.c.
\end{eqnarray}
Note that in this case, had we set $\phi_1=\phi_2$ the interaction
would vanish. Unlike the previous example for the case of an
interacting triplet with the higher spin $3$  with two scalars  one
cannot put the scalars $\phi_1$ and $\phi_2$ to be equal to each
other so one needs a complex scalar in analogy with scalar
electrodynamics. There is one more difference with respect to the
previous example, namely when doing the deformation to the
$AdS_{\cal D}$ case , apart from changing ordinary derivatives to
covariant ones, both the Lagrangian and gauge transformation rules
for scalars get deformed. Namely one has
\begin{eqnarray}
&&\Delta L_{free}=- \frac{1}{L^2} (6h_{\mu}^{\mu \rho} h_{\nu
\rho}^\nu - 48 h_\mu^{\mu \nu} D_\nu -({\cal D} -3) h_{\mu \nu
\rho} h^{\mu \nu \rho} + \nonumber \\
&&+18( {\cal D} +3) D^\mu D_\mu + (2{\cal D}-6)\ (
\phi_1^2+\phi_2^2) )
\end{eqnarray}
\begin{equation}\label{irrAdSV2}
\Delta L_{int} = i \ C_{3,0}\ \frac{6{\cal D}}{L^2}\ D^\mu \ (
\phi_1 \nabla_\mu \phi_2 - \phi_2 \nabla_\mu \phi_1)+ h.c.
\end{equation}
\begin{equation}\label{gtAdS}
\delta_{AdS} \phi_1 = \delta_0 \phi_1 - i \ C_{3,0}\
\frac{6}{L^2}\lambda^\mu_\mu \phi_2, \quad \delta_{AdS} \phi_2 =
\delta \phi_2 +i \ C_{3,0}\ \frac{6}{L^2}\lambda^\mu_\mu \phi_1,
\end{equation}
\begin{equation}
\delta_{AdS}C_{\mu \nu} = \delta C_{\mu \nu} + \frac{2(1-{\cal
D})}{L^2}\lambda_{\mu \nu}+ \frac{2}{L^2}g_{\mu \nu}
\lambda^\rho_\rho.
\end{equation}

\setcounter{equation}0\section{Irreducible
HS gauge field coupled to scalars in  AdS}\label{VAdS2}

In this section we will study the much simpler case of an
irreducible HS field in AdS coupled to two massive free scalars.
From the triplet we readily find the irreducible spin-$s$ gauge
field upon imposing the conditions \cite{Francia:2002pt}
\begin{equation}\label{fpd2}
{\cal W}_{p}=
\Lambda^{[p]}=0, \qquad p \geq 1\,,
\end{equation}
with the definitions (\ref{hatS}). Note that this corresponds to
setting the {\it compensator} field of \cite{Francia:2002pt}
 to zero. The corresponding Lagrangian
formulation which gives the equation (\ref{fpd2}) as a equation of
motion is available \cite{Buchbinder:2001bs}, \cite{Sagnotti:2003qa}
but since it is more complicated we add the equation (\ref{fpd2}) to
triplet ''by hand".
 The above conditions lead to traceless gauge parameters
and double-traceless gauge fields, namely ($'$ denotes the trace)
\begin{eqnarray}\label{irrcon}
&& \phi^{''}=0 \rightarrow (M_{33})^2 \ |\phi_3 \rangle=0 \,,\\
&&\lambda^{'}=0 \rightarrow M_{33}\ |\Lambda_3\rangle=0 \,.\nonumber
\end{eqnarray}
This simplifies the calculations since the only non-vanishing matrix
elements are now $X^1_{s-2p,0,p=0,1}$, $X^3_{31;s-2p-1,0,p=0}$ and
$X^2_{33;s-2p-2,0,p=0}$. The computation follows the steps of
(\ref{Es00A}) but in this case we keep only terms up to the first
power of $2M_{33}=X^2$ (see the Appendix E for details). The final
result is
\begin{eqnarray}\label{irrsol}
&&X^3_{31;s-1,0,0}=-X^3_{32;s-1,0,0}=s X^1_{s,0,0} \,,\\
&&X^2_{33,s-2,0,0}=-X^1_{s-2,0,1} + ({s-1\over 3} (2s^2+ (3{\cal
D} -4)s -6 )-2(s-2)) X^1_{s,0,0}.
\end{eqnarray}

Based on the solution above the gauge transformation is given by
the direct covariantization of  (\ref{FGT2}) for $p=0$. The cubic
interaction is given by
\begin{equation}\label{irrAdSV}
{\cal L}_{s00}= C_{s,0}\int  \ d^{\cal D} x \sqrt{g}\Biggl\{ \phi\cdot J^\nabla_{s}\ +
\left( {s-1 \over 6L^2} [2s^2 + (3{\cal D} -4)s-6]-
\frac{(s-2)}{L^2}\right)\phi^{'} \cdot J^\nabla_{s-2}  \Biggl\} \ + \ h.c.
\end{equation}
where the currents $J^\nabla_{s}$ and $J^\nabla_{s-2}$  are the AdS covariantizations
of the corresponding flat symmetric ones in (\ref{Jp}),
however only  the {\it double-traceless} part of $J_s$
and the {\it traceless} part of $J_{s-2}$ survive.
The interaction Lagrangian is a function of just one unknown normalization
constant $C_{s,0}$.

Let us now discuss our result (\ref{irrAdSV}). Firstly, the flat
space restriction of (\ref{irrAdSV}), (i.e. dropping the $1/L^2$
terms), implies that an irreducible spin-$s$ HS gauge field couples
to totally  symmetric, conserved currents. These currents coincide
with the ones obtained by Berends et. al. in \cite{Berends:1985xx}. For $s\geq 4$ only
the double-traceless part of $J_s$ survives.

In AdS, we define the modified current
\begin{equation}\label{AdSJ}
J^{AdS}_s= J^\nabla_s + \left({s-1 \over 6L^2} [2s^2 + (3{\cal D}
-4)s-6]-\frac{(s-2)}{L^2}\right) \ g \ J^\nabla_{s-2}\,,
\end{equation}
where $g$ is the AdS metric.  This current is not conserved but
satisfies \footnote{Remember the rule (\ref{irrcon}) applied to
our computation in the traceless case. This means we dropped all
terms proportional to any power of the metric g in the BRST
computation}
\begin{equation}\label{dtracediv}
[\ \nabla \cdot J^{AdS}_s\ ]^{traceless}=0\,.
\end{equation}
This condition implies that the
covariantized flat spin-$s$ current $J^\nabla_s$ fails to be conserved by order $1/L^2$ terms.

A few more comments are in order here.
In flat space, the conserved currents $J_s$ are not
 the only ones that couple consistently to
 irreducible HS gauge fields. They can be modified, at
 will,  by terms whose divergence gives zero upon contraction
 with the traceless gauge parameter $\lambda$.
 We fixed this freedom by imposing the
 conditions (\ref{fpd2}) i.e. setting  {\it compensator} field to zero.
 In \cite{Anselmi:1999bb}  an {\it additional}
 single zero trace  condition was used for the conserved
 currents in order to uniquely fix the form of the
 higher-spin currents in flat space.
 This condition was generalized in AdS by \cite{Manvelyan:2004mb}.
 In the above works, bulk conformal (or Weyl)
 invariance played a crucial role. We believe that
 our approach is more general since our HS gauge
 fields are coupled generically to massive scalars.
 It is also satisfying that our approach is
 naturally tied to BRST, as we believe that this is
 relevant for the application of our results to string theory.

\section{An alternative derivation of the cubic couplings}

In this section we present an alternative derivation of the cubic
interaction vertices of free massive scalars with HS triplets.
Although not yet under total control, the method is a generalization
of the standard Noether gauging in field theory and is based on a
surprisingly simple symmetry of the mass term in the free
Lagrangian. We present here explicitly the spin-2 and spin-3
coupling and leave the discussion of the symmetry and the spin-$s$
cases, with $s\geq 4$ for the future.

\subsection{Free massive scalars coupled to the spin-2 triplet}

Consider the Lagrangian for a massive free scalar field in flat
space
\begin{equation}
\label{freemassive} {\cal
L}=\frac{1}{2}(\partial^\mu\phi)(\partial_\mu\phi)
+\frac{1}{2}m^2\phi^2\,.
\end{equation}
The idea is that the spin-2 triplet will emerge through the gauging of a
symmetry of the above. The spin-2 triplet involves\footnote{In the section we
always solve the algebraic equation for the fields $C$.} the unconstrained
symmetric field $h_{\mu\nu}$, the auxiliary scalar $D$, while the
gauge parameter is the vector $\lambda_\mu$. We are looking for a transformation of the fields $\phi$ that induces a change of
(\ref{freemassive}) of the form
\begin{equation}
\delta{\cal L} =(\partial^\mu\lambda^\nu) T_{\mu\nu}
=\frac{1}{2}\delta h^{\mu\nu}T_{\mu\nu}\,.
\end{equation}
Had we found such a transformation, we would conclude; first
that $T_{\mu\nu}$ is a conserved current and second that the
interaction of the massive free scalars with the triplet is of the
form
\begin{equation}
{\cal L}_{int} =-c_1\frac{1}{2}h^{\mu\nu}T_{\mu\nu} +c_2(h'-2D)T\,.
\end{equation}
$c_1$ and $c_2$ are arbitrary constants. Notice that gauge
invariance cannot determine the "spin-0 current" $T$ since the
variation of the second term identically vanishes.

At this moment one might think we are just describing the gauging of
diffeomorphisms. Indeed, using
$\delta\phi=c_1\epsilon_\mu\partial_\mu\phi$ one finds the canonical
energy-momentum tensor
\begin{equation}
\label{spin2canonical} T_{\mu\nu}^{can}
=(\partial_\mu\phi)(\partial_\nu\phi)-\eta_{\mu\nu}{\cal L}\,.
\end{equation}
However, this is {\it not} what the BRST analysis gives, both for
the transformation of scalars and for the conserved spin-2 current
$T_{\mu\nu}$. Instead, we will look for a new principle that may fix
the scalar field transformations not only for the spin-2 case but
for HS as well.

To this effect, consider the most general infinitesimal (i.e. involving one derivative)
transformation of scalars with gauge parameter the vector
$\lambda_\mu$
\begin{equation}
\label{gaugetransf2} \delta\phi=c_1(\lambda^\mu\partial_\mu
+\kappa(\partial\cdot\lambda))\phi\,.
\end{equation}
Next we demand that (\ref{gaugetransf2}) leaves invariant the mass
term in (\ref{freemassive}) up to total derivatives (that do not play
a role in the action). This uniquely determines the parameter
$\kappa=1/2$ and reproduces the corresponding
transformation for scalars (\ref{gt21}) or (\ref{gt22}) (we can set $\phi_1=\phi_2$ there).
It is important to note that for $\kappa=1/2$
(\ref{gaugetransf2}) is {\it not} a Weyl transformation. Using
(\ref{gaugetransf2}) we can vary the kinetic term in
(\ref{freemassive}) and we straightforwardly obtain the conserved
current
\begin{equation}
\label{spin2current} T_{\mu\nu} =
\frac{1}{2}\left((\partial_\mu\phi)(\partial_\nu\phi)-\phi\partial_\mu\partial_\nu\phi\right)\,,
\end{equation}
which coincides with the corresponding conserved current in (\ref{li200}). It should be noted that (\ref{spin2current}) and the canonical
energy-momentum tensor (\ref{spin2canonical}) give rise to the same
conserved quantitites (energy and momentum) on-shell. Hence, we
expect that the transformation (\ref{gaugetransf2})  is actually
equivalent to diffeomorphisms, the additional term being equivalent
to a Field Redefinition in the BRST language.

Next we move to AdS. We covariantize the derivatives in the
transformation (\ref{gaugetransf2}) and we observe that it still
leaves the mass term invariant up to total derivatives. However, the
variation of the kinetic term is now different. Up to total
derivatives we obtain
\begin{equation}
\delta{\cal L} =
c_1\frac{1}{2}(\nabla^\mu\lambda^\nu)\left((\nabla_\mu\phi)(\nabla_\nu\phi)-\phi\nabla_{\mu}\nabla_\nu\phi
-\frac{{\cal D}-1}{4L^2}\eta_{\mu\nu}\phi^2\right)\,.
\end{equation}
Hence, in AdS the coupling is modified as
\begin{equation}
-c_1\frac{1}{2}h^{\mu\nu}T_{\mu\nu}\rightarrow
-c_1\left[\frac{1}{2}h^{\mu\nu}T_{\mu\nu}^{\nabla} -\frac{{\cal
D}-1}{8L^2}D\phi^2\right]\,,
\end{equation}
with $T_{\mu\nu}^{\nabla}$ denoting the covariantized current. This
is in agreement with the result (\ref{lintAdS200}).

\subsection{Free massive scalars coupled to the spin-3 triplet}

In this case we must have two different scalars to begin with. The
free Lagrangian is
\begin{equation}
\label{freemassive3} {\cal
L}=\frac{1}{2}(\partial^\mu\phi_i)(\partial_\mu\phi_i)
+\frac{1}{2}m^2\phi_i^2\,,\,\,\,\,\, i=1,2\,.
\end{equation}
The spin-3 triplet involves the symmetric tensor $h_{\mu\nu\rho}$,
the vector $D_\mu$, while the gauge parameter is the symmetric
tensor $\lambda_{\mu\nu}$. Hence, under the scalar field
transformation we expect that the free Lagrangian will vary as
\begin{equation}
\delta{\cal L} = q_1(\partial^\mu\lambda^{\nu\rho})T_{\mu\nu\rho}
+q_2(\partial^\mu\lambda')T_\mu\,,
\end{equation}
with $q_1$ and $q_2$ arbitrary constants. This would imply that the
coupling to the triplet is
\begin{equation}
\label{Lint3} {\cal L}_{int} =
-q_1\frac{1}{3}h^{\mu\nu\rho}T_{\mu\nu\rho} -q_2(h^{\mu
\nu}_{\nu}-2D_\mu)T_\mu\,.
\end{equation}
Notice the presence of the spin-1 conserved current $T_\mu$ and the
fact that $q_1$ and $q_2$ have different dimensions.

To construct the spin-3 current we seek first the most general
scalar field transformation that involves the gauge parameter
$\lambda_{\mu\nu}$ (and {\it not} its trace), that leaves the mass
term invariant. A simple calculation gives the result\footnote{Note
the similarity with  the generalized Lie derivative obtained in the
last of \cite{Bekaert:2005jf}.}
\begin{equation}
\label{gaugetransf3} \delta\phi_i
=q_1\left(\lambda^{\mu\nu}\partial_\mu\partial_\nu
+(\partial_\mu\lambda^{\mu\nu})\partial_\nu+
B(\partial_\mu\partial_\nu\lambda^{\mu\nu})\right)\phi_j\epsilon_{ij}\,,
\end{equation}
where we use the totally antisymmetric tensor $\epsilon_{ij}$,
$i,j=1,2$. We note that the parameter $B$ cannot be fixed  by
requiring invariance of the mass term. However, applying the
transformation (\ref{gaugetransf3}) to the kinetic term of (\ref{freemassive3}) we get
\begin{equation}
\delta{\cal L} = q_1(\partial^{\mu}\lambda^{\nu\rho})\left(
(2B-1)(\partial_\mu\partial_\nu\phi_i)(\partial_\rho\phi_j) +
B(\partial_{\mu}\phi_i)(\partial_\nu\partial_\rho\phi_j)+B(\partial_\mu\partial_\nu\partial_\rho\phi_i)
\phi_j\right)\epsilon_{ij}.
\end{equation}
Symmetrizing the current, in order to produce the $\delta
h^{\mu\nu\rho}T_{\mu\nu\rho}$ term, we find $B=1/4$, in agreement
with the corresponding scalar fields transformations (\ref{gt31})
and (\ref{gt32}).
  The conserved spin-3 current we
find is
\begin{equation}
T_{\mu\nu\rho}
=\frac{1}{4}\left((\partial_\mu\partial_\nu\partial_\rho\phi_i)\phi_j-
3(\partial_\mu\partial_\nu\phi_i)(\partial_\rho\phi_j)\right)\epsilon_{ij}\,,
\end{equation}
which coincides (up to an overall numerical factor) with (\ref{lint300}).

Passing on to AdS, we first notice that the covariantization  of the
transformation (\ref{gaugetransf3}) with $B=1/4$ leaves the mass
term invariant. However, the variation of the kinetic term is now
altered. Alter a lengthy calculation we find that in
AdS the coupling is modified as
\begin{equation}\label{300AdST}
-q_1\frac{1}{3}h^{\mu\nu\rho} T_{\mu\nu\rho} \rightarrow
-q_1\left[\frac{1}{3}h^{\mu\nu\rho}T_{\mu\nu\rho}^{\nabla}
-\frac{1}{2L^2}\left(\frac{2}{3}h^{\mu\nu}_{\,\,\,\nu}-({\cal
D}+1)D^\mu\right)T_{\mu}^\nabla\right]\,,
\end{equation}
where the covariantized spin-1 current is
\begin{equation}
T_\mu^\nabla=(\nabla_\mu\phi_i)\phi_j\epsilon_{ij}\,.
\end{equation}
This result is in agreement with the corresponding result of the
BRST analysis (\ref{irrAdSV2}) for $s=3$. Indeed, a piece of
(\ref{300AdST}) proportional to $h^\mu -2D^\mu$ can be associated to a modification of
the gauge transformation as in (\ref{gtAdS}) and the remaining can be
 seen as a modification of the coupling. Nevertheless, a highly
 non-trivial check of (\ref{300AdST}) is that when we set $h'_\mu=2D_\mu$ we
  get the result (\ref{AdSJ}) which was gotten by a totally independent method.
  The term involving $q_2$ in the interaction
(\ref{Lint3}) is simply modified by $T_\mu\rightarrow
T_\mu^\nabla$.

\section{Summary and Outlook}

We have applied the general BRST procedure of
 \cite{Buchbinder:2006eq} to construct the cubic
 interaction Lagrangian vertex of HS triplets
 coupled to free massive scalars. Although this is
 the simplest possible case of HS interactions, it still
 involves considerable technical tasks. We were able to
give closed expressions for the  vertex in flat and AdS spaces,
however, the AdS expressions are still quite involved.

The cubic vertex in flat space has an interesting structure. Namely,
the spin-$s,s-2,...,0/1$ modes that are propagated by a spin-$s$
triplet couple independently to corresponding conserved currents
constructed from the scalars. In flat space these are the currents
constructed long ago by Berends et. al. \cite{Berends:1985xx}. In
AdS, the situation changes and generically both the gauge variations
and the couplings are deformed by $1/L^2$ corrections. Although
there might be a pattern for the AdS deformation we were not able to
find it. We can pass to irreducible HS modes by simply setting the
{\it compensator} fields \cite{Francia:2002pt} to zero. This gauge
choice allows us to use the same symmetric and conserved currents
found above for the coupling of scalars to irreducible HS gauge
fields in flat space. Again, in AdS the currents are deformed by
$1/L^2$ terms. We never use conformal or Weyl invariance in our
construction as in the works \cite{Anselmi:1999bb,
Manvelyan:2004mb}. The detailed expressions for the spin-2 and
spin-3 cases are given. The latter results are reproduced by an
alternative method based on the idea that HS gauge fields arise via
the gauging of higher-derivative symmetries of free matter
Lagrangians.

There are many interesting applications and extensions of our work.
Since we were able to couple HS gauge fields to massive scalars our
results can be readily used in holography. In particular, an obvious
implication of our results is the existence of an infinite set of
Ward identities associated to composite scalar operators in
conformal field theories dual to HS gauge theories\footnote{Similar
ideas were discussed in \cite{Mikhailov:2002bp}.} Also, our
calculations are the first step towards the construction of the
Lagrangian cubic vertex of HS gauge fields with spins $s\neq 0$ in
AdS. The holographic interpretation of such a calculation will give
the three-point functions of the energy momentum tensor and of an
infinite set of higher spin conserved currents in the boundary CFT.
This way we hope to understand the meaning of the parameters present
in three-point functions of conserved currents of generic CFTs
\cite{OP}. These issues will be studied in a forthcoming work.

Finally, a few words are reserved for
the alternative derivation of the cubic couplings.
A scalar field deformation in terms of a
 vector-like gauge parameter $\lambda_\mu$ is simply
associated to diffeomorphisms. It would be extremely interesting to
understand the origin of our higher-derivative scalar
transformations, those that involve tensor gauge parameters, that
keep the mass term invariant. It is conceivable that they indicate a
broader structure for the underlying "spacetime", perhaps one that
involves tensor coordinates. It would also be interesting to study
the algebraic structure of our higher-derivative transformations. We
intent to come back to these exciting questions in the near future.

 \vspace{1cm}

\noindent {\bf Acknowledgments.} We`are grateful to X. Bekaert, N.
Boulanger, O. Hohm and M. Vasiliev for useful comments.
 The work of A.F. is partially
supported by the European Commission, under RTN program
MRTN-CT-2004-0051004 and by the Italian MIUR under the contract PRIN
2005023102.
The work of N. I. was partially supported by the "France-Greece Common
 Research Program in Science and Technology", K.A. 2342. The work of A. C. P. was
partially supported by the PYTHAGORAS II Research Program K.A.2101,
of the Greek Ministry of Higher Education. The work of M.T. was
supported by the European Commission under RTN program
MRTN-CT-2004-512194.


\renewcommand{\thesection}{A}

\setcounter{equation}{0}

\renewcommand{\theequation}{A.\arabic{equation}}
\section{Basic Definitions}\label{ApA}

\subsection{Definition of $|V\rangle$ and $|W \rangle$}

We define two linearly independent combinations of variables with
ghost number zero
\begin{equation}\label{Defab}
\gamma^{ij,+}=c^{i,+} b^{j,+}, \ \ \ \ \beta^{ij,+}=c^{i,+} b^j_0\,.
\end{equation}
Then,  the most general expansion of the vertex in terms of ghost
variables has the form
\begin{eqnarray}\label{VExp}
&|V\rangle= \Bigl\{X^1 + X^2_{ij} \gamma^{ij,+}+ X^3_{ij}
\beta^{ij,+} + X^4_{(ij);(kl)}\gamma^{ij,+}\gamma^{kl,+}+
X^5_{ij;kl}\gamma^{ij,+}\beta^{kl,+}+ \nonumber \\
&+X^6_{(ij);(kl)}\beta^{ij,+}\beta^{kl,+}
+X^7_{(ij);(kl);(mn)}\gamma^{ij,+}\gamma^{kl,+}\gamma^{mn,+}+
X^8_{(ij);(kl);mn}\gamma^{ij,+}\gamma^{kl,+}\beta^{mn,+}+ \nonumber \\
&+X^9_{ij;(kl);(mn)}\gamma^{ij,+}\beta^{kl,+}\beta^{mn,+}+
X^{10}_{(ij);(kl);(mn)}\beta^{ij,+}\beta^{kl,+}\beta^{mn,+}\Bigl\}
|-\rangle_{123}\ .
\end{eqnarray}
 The coefficients $X^l$ depend only on operators $\alpha^{i +}$ and
$p^i$, which means that they can be written as
\begin{eqnarray}\label{XEXP}
&X^{l}_{(\dots)}=
X^{l}_{n_1,n_2,n_3;m_1,k_1,m_2,k_2,m_3,k_3;p_1,p_2,p_3,r_{12},r_{13},r_{23}(\dots)}
\nonumber \\
&(l_0^1)^{n_1} \dots (l^{+,1})^{m_1} (I^{+,1})^{k_1} \dots
(M^{+,11})^{p_1}\dots (M^{+,12})^{r_{12}} \dots
\end{eqnarray}
where
\begin{equation}\label{LIVGen}
l_0^{ij}=(l_0^{11},l_0^{22},l_0^{33})=(l_0^{1},l_0^{2},l_0^{3})
\quad  l^{ij}= (l^{1},I^{1},l^{2},I^{2},l^{3},I^{3}),
\end{equation}
\begin{equation}
  I^{1}=\alpha^{ \mu, 1}(p_{\mu}^2-p_{\mu}^3) , \quad
I^{2}=\alpha^{ \mu, 2}(p_{\mu}^3-p_{\mu}^1), \quad I^{3}=\alpha^{
\mu, 3}(p_{\mu}^1-p_{\mu}^2),
\end{equation}
\begin{equation}
l^{i}=l^{ii}, \quad M^{ij} =\frac{1}{2}\alpha^{i \mu} \alpha^j_\nu
\end{equation}
In a similar manner one has  the following expansion for the
operator $|W\rangle$
\begin{eqnarray}\label{WExp}
&|W\rangle= \Bigl\{W_i^1b^{i,+} + W_i^2 b^i_{0}+ W^3_{i;jk}b^{i,+}
\gamma^{jk,+}+ W^4_{i;jk}b^{i,+} \beta^{jk,+}+ W^5_{i;jk}b^i_{0}
\beta^{jk,+}+
  \nonumber \\
&W^6_{i;(jk);(lm)}b^{i,+}\gamma^{jk,+}\gamma^{lm,+}+W^7_{i;jk;lm}b^{i,+}\gamma^{jk,+}\beta^{lm,+}+
W^8_{i;(jk);(lm)}b^{i,+}\beta^{jk,+}\beta^{lm,+}+ \nonumber \\
&W^9_{i;(jk);(lm)}b^i_{0}\beta^{jk,+}\beta^{lm,+}
+W^{10}_{i;(jk);(lm);pn}b^{i,+}\gamma^{jk,+}\gamma^{lm,+}\beta^{pn,+}
+
\nonumber \\
&W^{11}_{i;jk;(lm);(pn)}b^{i,+}\gamma^{jk,+}\beta^{lm,+}\beta^{pn,+}
+
W^{12}_{i;(jk);(lm);(pn)}b^{i,+}\beta^{jk,+}\beta^{lm,+}\beta^{pn,+}\Bigl\}|-\rangle_{123}\,.
\end{eqnarray}
Alternatively, one can expand in terms of the operators $l^{ij+}=
\alpha^{\mu i, +} p^j_{\mu}$ and $l_0^{ij}=p^{\mu i}p_{\mu}^j$ but
one has to bear in mind that not all of these are independent due to
the momentum conservation law $p^1_\mu + p^2_\mu + p^3_\mu=0$ (see
\cite{Buchbinder:2006eq} for more details).

\renewcommand{\thesection}{B}

\setcounter{equation}{0}

\renewcommand{\theequation}{B.\arabic{equation}}

\subsection{Definition of  the F, G  functions and their properties}

The following identities hold:
\begin{eqnarray}\label{id0}
&& \sum_{u=0}^{[{n-1 \over 2}]} \ \bn{n}{2u+1} \ a^{n-2u} \
x^{2u}= {a \over
2x} \ [(a+x)^n-(a-x)^n] \\
&& \sum_{u=0}^{[{n \over 2}]}\  \bn{n}{2u} \ a^{n-2u} \ x^{2u}= {1
\over 2} \ [(a+x)^n+(a-x)^n] \nonumber \\
&& \sum_{k=0}^{n} a^k (a+x)^{n-k}= -{1 \over x}
(a^{n+1}-(x+a)^{n+1}) \nonumber
\end{eqnarray}

Then we define
\begin{eqnarray}\label{F0}
&& F(n,0,Y,X)= \sum_{k=0}^{n} \sum_{u=0}^{[{n-k\over 2}]} \
\bn{n-k+1}{2u+1} \ Y^{n-2u+1} \ X^{2u}= \nonumber \\
&&={1\over 2}\sum_{k=0}^{n+1} \ \bn{n+2}{k} \ Y^{k+1} \ X^{n-k} \
(1+(-1)^{n-k})=
\\
&&={Y \over 2X^2} \ [-2 Y^{n+2} +(Y+X)^{n+2}+(Y-X)^{n+2}]
\nonumber
\end{eqnarray}
and

\begin{eqnarray}\label{F1}
&& F(n,1,Y,X)= \sum_{k=0}^{n} \sum_{u=0}^{[{n-k \over 2}]} \
\bn{n-k}{2u} \ Y^{n-2u} \ X^{2u}= \nonumber \\
&&={1\over 2}\sum_{k=0}^{n+1} \bn{n+1}{k} \ Y^{k} \ X^{n-k} \
(1+(-1)^{n-k})=
\\
&&={1 \over 2X} \ [(Y+X)^{n+1}-(Y-X)^{n+1}] \nonumber
\end{eqnarray}
The function $F(n,\lambda,Y,X), \ \lambda=0,1$ as defined above has
the property that it is an expansion in even powers of $X$.

Using (\ref{id0}) we can show the following identities:
\begin{eqnarray}\label{Fid}
&&\sum_{k=0}^n \ Y^k\ F(n-k,0,Y,X)= {Y \over X^2} \ [-(n+3)Y^{n+2}
+ F(n+2,1,Y,X)] \nonumber \\
&&\sum_{k=0}^n \ Y^k\ F(n-k,1,Y,X)= {1\over Y}F(n,0,Y,X) \nonumber
\\
&&\partial_Y F(n,0,Y,X)= {1 \over Y} F(n,0,Y,X) + (n+2)
F(n-1,0,Y,X) \\
&&\partial_Y F(n,1,Y,X)= (n+1)F(n-1,1,Y,X) \nonumber\\
&& \partial_{X^2} F(n,0,Y,X)= -{1 \over X^2} F(n,0,Y,X) + {Y \over
2X^2} F(n,1,Y,X) \nonumber \\
&&\partial_{X^2} F(n,1,Y,X)= -{1 \over 2X^2} F(n,1,Y,X) + (n+1)
{Y^n \over 2X^2} + {n+1 \over 2Y}F(n-2,0,Y,X) \nonumber
\end{eqnarray}
All of the above identities do not produce any negative powers of
$Y$ or $X^2$. This will be useful in Appendix \ref{ApC}.

We finally define the following functions:
\begin{eqnarray}\label{G}
&&G_e(n,0,Y,X)= \sum_{k=0}^{n} \sum_{u=0}^{[{n-k\over 2}]}\
\bn{n-k}{
2u}\ F(n-2u,0,Y,X) X^{2u}= \nonumber \\
&& ={Y \over 2X^2} [-2 Y^2 \ F(n,1,Y,X)+(Y+X)^2\ F(n,1,Y+X,X)
\nonumber \\
&&+(Y-X)^2\ F(n,1,Y-X,X)] \nonumber \\
&&G_e(n,1,Y,X)= \sum_{k=0}^{n} \sum_{u=0}^{[{n-k\over 2}]}\
\bn{n-k}{
2u}\ F(n-2u,1,Y,X) X^{2u}= \nonumber \\
&& ={1 \over 2X} [(Y+X)\ F(n,1,Y+X,X) \nonumber \\
&&-(Y-X)\ F(n,1,Y-X,X)]\\
&&G_o(n,0,Y,X)= \sum_{k=0}^{n} \sum_{u=0}^{[{n-k\over 2}]}\
\bn{n-k+1}{2u+1}\ F(n-2u,0,Y,X) X^{2u}= \nonumber \\
&& ={Y \over 2X^2} [-2 Y \ F(n,0,Y,X)  \nonumber \\
&&+(Y+X)\
F(n,0,Y+X,X)+(Y-X)\ F(n,0,Y-X,X)] \nonumber \\
&&G_0(n,1,Y,X)= \sum_{k=0}^{n} \sum_{u=0}^{[{n-k\over 2}]}\
\bn{n-k+1}{2u+1}\ F(n-2u,1,Y,X) X^{2u} \nonumber \\
&& ={1 \over 2X} [(F(n,1,Y+X,X)- F(n,1,Y-X,X)]\nonumber
\end{eqnarray}
The functions with arguments $Y+X$ and $Y$ are related. Using:
\begin{eqnarray}\label{Fid2}
&&(Y+X)^{n+2}= {X^2 \over Y} \ F(n-1,0,Y,X) + X \ F(n,1,Y,X) +
Y^{n+1} \\
&&(Y+X)^{n+2}= {X^2 \over Y} \ F(n-1,0,Y,X) - X \ F(n,1,Y,X) +
Y^{n+1} \nonumber
\end{eqnarray}
one can write i.e.
\begin{eqnarray}\label{Fid3}
&&F(n,1,Y \pm X,X)= \pm {1 \over 2X} \ [ (Y\pm 2X)^{n+1}-  Y^{n+1}] = \\
&&= \pm {2X \over Y} \ F(n-1,0,Y,2X) + F(n,1,Y,2X) \nonumber
\end{eqnarray}
\begin{eqnarray}\label{Fid5}
&&F(n,0,Y\pm X,X)= {Y \pm X \over 2X^2} \ [ -2(Y \pm X)^{n+2}
\nonumber \\
&&+(Y \pm 2X)^{n+2}+ Y^{n+2}] =
 {Y \pm X \over X} \ F(n,0,Y,2X).
\end{eqnarray}
Using all of the above we can easily show for example that
\begin{eqnarray}\label{Fid6}
&&G_e(n,0,Y,X)=  {Y \over 2X^2} \ [-2Y^2 \ F(n,1,Y,X) \nonumber \\
&&+
 \ 4X^2 \
F(n-1,0,Y,2X) + \ (Y^2+X^2)\ F(n,1,Y,2X)] \nonumber
\end{eqnarray}
which makes it easy to expand in a single series expansion in terms
of $Y$ and $X$ using (\ref{F0},\ref{F1}). Finally we will define the
following compact expression which will make  the presentation of
our results in the main text easier
\begin{eqnarray}\label{Fid7}
&&{\tilde G}_e(n,\lambda, Y, X; a) = \sum_{k=0}^{n}
\sum_{u=0}^{[{n-k\over 2}]}\ \bn{n-k}{
2u}\ F(n-a-2u,\lambda,Y,X) X^{2u}=  \nonumber  \\
&&=\ G_e(n-a,\lambda, Y, X) + \ \sum_{u=0}^{[{n-a\over 2}]}\
\sum_{k=0}^{a-1}\bn{n-k}{ 2u}\ F(n-a-2u,\lambda,Y,X) X^{2u}
\nonumber\\
\end{eqnarray}
and
\begin{eqnarray}\label{Fid8}
&&{\tilde G}_o(n,\lambda, Y, X; a) = \sum_{k=0}^{n}
\sum_{u=0}^{[{n-k\over 2}]}\ \bn{n-k+1}{
2u+1}\ F(n-a-2u,\lambda,Y,X) X^{2u} = \nonumber  \\
&&=\ G_o(n-a,\lambda, Y, X) + \ \sum_{u=0}^{[{n-a\over 2}]}\
\sum_{k=0}^{a-1}\bn{n-k+1}{ 2u+1}\ F(n-a-2u,\lambda,Y,X) X^{2u}
\nonumber\\
\end{eqnarray}

\renewcommand{\thesection}{C}

 \setcounter{equation}{0}

\renewcommand{\theequation}{C.\arabic{equation}}

\section{Commutation relations}\label{ApB}

We wish to compute the commutators of $p^{\mu, i}, \ i=1,2$ and
$\alpha^{+,3}_\mu$ with strings of operators involving $l^{3i}$ and
$M^{33}$. We will use the following equation for a tensor $T_{\rho,
\dots}$:
\begin{eqnarray}\label{Dcom}
&&D^{\mu \nu}T_{\rho, \dots}=[p_{\mu}, p_\nu]T_{\rho, \dots}=-
{1\over L^2}\ ( g_{\nu \rho} T_{\mu \dots}-g_{\mu \rho} T_{\nu
\dots}) + \dots
\end{eqnarray}
We will set $L^2=1$ and will reinstate it only at the end of our
calculations based on dimensional analysis.

We start first with the momenta commutators. We drop the Fock index
from the oscillators. The following computation holds:
\begin{eqnarray}\label{S1}
&&S_{\nu}(n)=[\alpha^\mu  D^2_{\mu \nu},
(l_{32})^n]= \\
&&=\alpha^\mu \ \sum_{k=0}^{n-1}\ (l_{32})^k \ (\alpha_\nu
p_{2,\mu}-\alpha_\mu p_{2,\nu}) \ (l_{32})^{n-k-1}\nonumber
\end{eqnarray}

We can then show by induction that:
\begin{eqnarray}\label{S2}
&&S_{\nu}(n) = -2(1+\Theta_{n-2}) M_{33} (l_{32})^{n-1} p_{2, \nu}
+ n \
\alpha_\nu(l_{32})^{n} + \nonumber \\
&&+2M_{33} \sum_{k=0}^{n-2}\ \sum_{u=k}^{n-k-2} \ (l_{32})^{k+u} \
S_\nu(n-k-u-2)
\end{eqnarray}
Some straightforward manipulations show that
\begin{eqnarray}\label{S3}
&& S_{\nu}(1)= \alpha_\nu l_{32}-2 M_{33} p_{2, \nu} \\
&&S_{\nu}(2)- l_{32} S_{\nu}(1)= \alpha_\nu (l_{32})^{2}-2 M_{33}
p_{2, \nu} \nonumber \\ &&S_{\nu}(n)- l_{32} S_{\nu}(n-1)=
\alpha_\nu (l_{32})^{n} + 2 M_{33}\ \sum_{k=0}^{n-2}\ (l_{32})^{k}
\ S_\nu(n-k-2) \ , \  n \geq 3 \nonumber
\end{eqnarray}
where $\Theta_{n}$ is $0$ for $n<0$ and $1$ otherwise.

Then by algebraic manipulations, induction and  use of the formula:
\begin{eqnarray}\label{MasterF}
\sum_{k=0}^{m} \ \bn{n+k}{n}= \bn{n+m+1}{n+1}
\end{eqnarray}
we arrive at the following solution:
\begin{eqnarray}\label{S4}
&&S_{\nu}(n)=\\
&&\alpha_\nu \ \sum_{k=0}^{[{n-1 \over 2}]} \ \bn{n}{ 2k+1} \
(l_{32})^{n-2k} (M_{33})^k + \sum_{k=0}^{[{n \over 2}]} \
\bn{n}{2k} \ (l_{32})^{n-2k}
(M_{33})^k \ S_{\nu}(0) \nonumber \\
&& - \sum_{k=0}^{[{n-1 \over 2}]} \ \bn{n-1} {2k} \
(l_{32})^{n-1-2k} (M_{33})^{k+1} \ p_{2,\nu} \nonumber \\
&&-
\sum_{k=0}^{[{n-2 \over 2}]} \ \bn{n-2} {2k} \ (l_{32})^{n-1-2k}
(M_{33})^{k+1} \ p_{2,\nu} \nonumber
\end{eqnarray}
It is easy to deduce that:
\begin{eqnarray}\label{pcom}
&&[p_{2\mu}, (l_{32})^n]= - \sum_{k=0}^{n-1} \ (l_{32})^k \
S_{\nu}(n-k-1)
\end{eqnarray}
Then using (\ref{S4}) and (\ref{F0}), (\ref{F1}) we arrive at the
following relation:
\begin{eqnarray}\label{pcom2}
&&[p_{2\mu}, (l_{32})^n]= -{1\over L^2}\ [ F(n-2,0,Y,{X \over L})
\ \alpha_\mu - F(n-2,1,Y,{X \over L}) \ X^2 \ p_{2,\mu} \\
&&-F(n-3,1,Y,{X \over L}) \ Y \ X^2 \ p_{2,\mu} ] - F(n-1,1,Y,{X
\over L}) \ S_\nu (0) \nonumber
\end{eqnarray}
where we have reinstated the units $L$ and also we have set
$Y=l_{32}$ and $X^2=2 M_{33}$. Note that although naively it might
seem that $X$ can appear in odd powers, therefore making no sense,
actually as we mentioned in Appendix \ref{ApA} the function $F(n,
\lambda,Y,X))$  always  has an even argument in the variable $X$.

In a similar manner we can show that:
\begin{eqnarray}\label{acom}
&&[(l_{32})^n, \alpha^+_\mu]= n (l_{32})^{n-1} p_{2,\mu}- {1 \over
L^2} \ [ {L^2 Y\over X^2} \ (-n
\ Y^{n-1} + F(n-1,1,Y,{X \over L}) \ ) \ \alpha_\mu  \nonumber \\
&& -( {1\over Y} \ F(n-3,0,Y,{X \over L})+ F(n-4,0,Y,{X \over L}))
\ X^2\ p_{2, \mu} \nonumber \\
&&  + {L^2 \over Y}\ F(n-2,0,Y,{X \over L})\ S_{\mu}(0)]
\end{eqnarray}
Finally we can compute the commutators of momenta and  oscillators
with $F(n, \lambda, Y, X)$:
\begin{eqnarray}\label{pcF0}
&&[p_\mu, F(n,0,Y,{X \over L})]= -{1\over L^2} \ \{ \
\tilde{G}_o(n,0,Y,{X
\over L};1)\ \alpha_\mu - \\
&&(\tilde{G}_o(n,1,Y,{X \over L};1) + Y \tilde{G}_o(n,1,Y,{X \over
L};2)) \ X^2 \ p_\mu + L^2 \tilde{G}_o(n,1,Y,{X \over L};0)\
S_\mu(0) \nonumber
\end{eqnarray}
\begin{eqnarray}\label{pcF1}
&&[p_\mu, F(n,1,Y,{X \over L})]= -{1\over L^2} \{ \
\tilde{G}_e(n,0,Y,{X
\over L};2)\ \alpha_\mu - \\
&&(\tilde{G}_e(n,1,Y,{X \over L};2) + Y \tilde{G}_e(n,1,Y,{X \over
L};3)  )\ X^2 \ p_\mu + L^2 \tilde{G}_e(n,1,Y,{X \over L};1)\
S_\mu(0)\} \nonumber
\end{eqnarray}

\begin{eqnarray}\label{acF0} \nonumber
&&[F(n,0,Y,{X \over L}), \alpha^{+}_\mu]= (\partial_Y F(n,0,Y,{X
\over L})\ (p_\mu+ {Y\over  X^2} \alpha_\mu) + 2
\partial_{X^2}F(n,0,Y,{X \over L}) \ \alpha_\mu \\
&& -{1\over L^2} \ \{ \ {L^2 Y\over X^2} \tilde{G}_o(n,1,Y,{X \over
L};0)\ \alpha_\mu -({1\over Y}\tilde{G}_o(n,0,Y,{X \over L};2)
 \\
&&+ \tilde{G}_o(n,0,Y,{X \over L};3))\ X^2 \ p_\mu +  {L^2\over Y}
\tilde{G}_o(n,0,Y,{X \over L};1)\ S_\mu(0) \} \nonumber
\end{eqnarray}

\begin{eqnarray}\label{acF1} \nonumber
&&[F(n,1,Y,{X \over L}), \alpha^{+}_\mu]= (\partial_Y F(n,1,Y,{X
\over L})\ (p_\mu+ {Y\over  X^2} \alpha_\mu) + 2
\partial_{X^2}F(n,1,Y,{X \over L}) \ \alpha_\mu \\
&& -{1\over L^2} \ \{ \ {YL^2 \over X^2} \tilde{G}_e(n,1,Y,{X \over
L};1)\ \alpha_\mu -({1\over Y}\tilde{G}_e(n,0,Y,{X \over L};3)
 \\
&&+ \tilde{G}_e(n,0,Y,{X \over L};4))\ X^2 \ p_\mu +
 {L^2\over Y} \tilde{G}_e(n,0,Y,{X \over L};2)\ S_\mu(0) \}
\nonumber
\end{eqnarray}

Finally we can easily show:
\begin{eqnarray}\label{acomX}
&&[X^{2p}, \alpha^+_\mu]= 2p X^{2(p-1)} \alpha_\mu.
\end{eqnarray}

This completes all the possible commutators needed for the
computations of the next Appendix.

\renewcommand{\thesection}{D}

\setcounter{equation}{0}

\renewcommand{\theequation}{D.\arabic{equation}}

\section{Equations for $I^3$}\label{ApC}
In a similar manner we can work in the $I_3, \ l_{33}$ basis. We
define:
\begin{eqnarray}\label{defPsi}
&&a^\mu \ (D^1_{\mu\nu} + D^2_{\mu \nu})\ (I_3)^n= \Sigma_\nu(n) \\
&&a^\mu \ (D^1_{\mu\nu} - D^2_{\mu \nu})\ (I_3)^n= \Psi_\nu(n)
\nonumber
\end{eqnarray}

By induction we can show:
\begin{eqnarray}\label{ppcomI}
&&[p_{1\mu}+p_{2\mu}, (I_{3})^n]= -{1\over L^2}\ [ {1\over Y}
F(n-2,0,Y,{X \over L}) \ (l_{31}+l_{32}) \ \alpha_\mu \\
&& - (F(n-2,1,Y,{X \over L}) + Y F(n-3,1,Y,{X \over L})) \ X^2 \
(p_{1,\mu} +p_{2,\mu}) \ ] \nonumber \\
&&- F(n-1,1,Y,{X \over L}) \
\Psi_\nu (0) \nonumber
\end{eqnarray}
where $Y=I_{3}$. We can also show that
\begin{eqnarray}\label{pmcomI}
&&[p_{1\mu}-p_{2\mu}, (I_{3})^n]= -{1\over L^2}\ [  F(n-2,0,Y,{X
\over L})  \ \alpha_\mu  \\
&&- (F(n-2,1,Y,{X \over L})+ Y F(n-3,1,Y,{X \over L})) \ X^2 \
(p_{1,\mu} -p_{2,\mu}) \ ] \nonumber \\
&& - F(n-1,1,Y,{X \over L}) \ \Sigma_\nu (0)\nonumber
\end{eqnarray}
\begin{eqnarray}\label{acomI}
&&[(I_{3})^n, \alpha^+_\mu]= n (I_{3})^{n-1}
(p_{1,\mu}-p_{2,\mu})- {1 \over L^2} \ [ {YL^2\over X^2} \ (-n
\ Y^{n-1} + F(n-1,1,Y,{X \over L}) \ ) \ \alpha_\mu \nonumber \\
&& -({1\over Y} \ F(n-3,0,Y,{X \over L}) + F(n-4,0,Y,{X \over L})
)\ X^2\ (p_{1,\mu}-p_{2,\mu}) \nonumber \\
&&+ {L^2 \over Y}\ F(n-2,0,Y,{X \over L})\
\Sigma_{\mu}(0)]\nonumber
\end{eqnarray}
Actually it is fairly easy to deduce the equivalents of (\ref{pcF0},
\ref{pcF1}, \ref{acF0}, \ref{acF1}) for the $I_3$. For example the
commutators of $(p_{1,\mu}-p_{2,\mu})$ with $F(n, \lambda, I_3, X)$
are deduced from (\ref{pcF0}, \ref{pcF1}) by substituting $p_{i,
\mu} \to (p_{1,\mu}-p_{2,\mu})$ and $S_\mu(0) \to \Sigma_\mu(0)$.
The same for the commutator of $\alpha^+_\mu$.

\renewcommand{\thesection}{E}

\setcounter{equation}{0}

\renewcommand{\theequation}{E.\arabic{equation}}

\section{The vertex for an irreducible HS gauge field}

In this Appendix we present the explicit computation for the BRST
equations for the cubic vertex for an irreducible HS field. In order
to compute (\ref{Es00A}) we need
\begin{eqnarray}\label{EIs00A}
&& _{123}\langle -| \ \sum_{n+2p=s-2}\ (I_3)^n \ (M_{33})^p \
l_{33}\
X^2_{33;p}= \\
&&_{123}\langle -|\ - \{(-Y^{s-2} + Y^{s-4} X^2 \bn{s-4}{s-6} \ )
X^2_{33;0} + O(X^4) \} \times (p_{1,\mu}+p_{2,\mu})\alpha_3^\mu
\nonumber
\end{eqnarray}
\begin{eqnarray}\label{EIs00A2}
&&_{123}\langle -| \ \sum_{n+2p=s}\ (I_3)^n \ (M_{33})^p \
l^+_{33}
X^1_{p}= \\
&&_{123}\langle -|- \{ \sum_{p=0}^1\ ( \ (s-2p) Y^{s-2p-1} \nonumber
\\
&&+ 3 (s-2p-2)^2 \ Y^{s-2p-3} X^2 ] {X^{2p} \over 2^p} \ X^1_{p}
\}\times (l_0^{11}-l_0^{22})  \nonumber \\
&&+ \{ \sum_{p=0}^1 \ p \ (Y^{s-2p}- Y^{s-2p-2} X^2
\bn{s-2p-2}{s-2p-4})\
{X^{2(p-1)} \over 2^{(p-1)}} \ X^1_{p} \nonumber \\
&& \sum_{p=0}^1 \ ( -(s-2p-1+{\cal D})\bn{s-2p}{s-2p-2}
-({\cal D}-1)\bn{s-2p}{s-2p-2} \nonumber \\
&&-\bn{s-2p}{s-2p-3}+ 2\bn{s-2p-1}{s-2p-2} \nonumber \\
&&+2
\bn{s-2p-2}{s-2p-3} )
Y^{s-2p-2}{X^{2p} \over 2^p} \ X^1_{s-2p,0,p} \nonumber \\
&&+ ( -(s-1+{\cal D}) \bn{s}{s-4} -({\cal D}-1)\bn{s}{s-4}-\bn{s}{s-5} \nonumber\\
&&-(3D-7)(\bn{s-1}{s-4}+ \bn{s-2}{s-5}) \nonumber \\
 &&-4(
\bn{s-1}{s-5} + \bn{s-2}{s-6}) Y^{s-4} X^2 \ X^1_0 \}\times
(p_{1,\mu}+p_{2,\mu})\alpha_3^\mu \nonumber
\end{eqnarray}
Finally plugging the expressions above in (\ref{Es00A}) and making
use of (\ref{irrcon}) we arrive at (\ref{irrsol}).

\end{document}